\documentclass[prl,aps,amssymb,twocolumn,showpacs,superscriptaddress]{revtex4}
\usepackage{graphicx}

\begin{document}

\title{Environmental effects in the third moment of voltage fluctuations in a tunnel junction} 

\author{B.~Reulet}
\affiliation{Departments of Applied Physics and Physics, Yale University, New Haven CT 06520-8284, USA} 
\affiliation{Laboratoire de Physique des Solides, UMR8502, b\^at 510, Universit\'e Paris-Sud 91405 Orsay, France}
\author{J. Senzier}
\author{D.E.~Prober} 
\affiliation{Departments of Applied Physics and Physics, Yale University, New Haven CT 06520-8284, USA}

\date{\today} 
\begin{abstract}

We present the first measurements of the third moment of the voltage fluctuations in a conductor. 
This technique can provide new and complementary information on the electronic transport in conducting systems. The measurement was performed on non-superconducting tunnel junctions as a function of voltage bias, for various temperatures and bandwidths up to 1GHz. The data demonstrate the significant effect of the electromagnetic environment of the sample.
\end{abstract}
\maketitle

Transport studies provide a powerful tool for investigating electronic properties of a conductor.  The $I(V)$ characteristic (or the differential resistance $R_{diff}=dV/dI$) contains partial information on the mechanisms responsible for conduction.  A much more complete description of transport in the steady state, and further information on the conduction mechanisms, is given by the probability distribution of the current $P$, which describes both dc current $I(V)$ and the fluctuations. Indeed, even with a fixed voltage $V$ applied, $I(t)$ fluctuates, due to  
the discreteness of the charge carriers, the probabilistic character of scattering and the fluctuations of the population of energy levels at finite temperature $T$ \cite{BuBlan}.

The current fluctuations are characterized by the moments of the probability distribution $P$ of order two and higher. Experimentally, the average over $P$ is obtained by time averaging. Thus the average current is the dc current $I=\left<I(t)\right>$, where $\left<.\right>$ denotes time average. The second moment (the variance) of $P$, $\left<\delta I^2\right>$, measures the amplitude of the current fluctuations, with $\delta I(t)=I(t)-I$. The third moment $\left<\delta I^3\right>$ (the skewness) measures the asymmetry of the fluctuations \cite{cumulant}. The existence of the third moment is related to the breaking of time reversal symmetry by the dc current; at zero bias, $I=0$ and positive and negative current fluctuations are equivalent, so $\left<\delta I^3\right>=0$.
An intense theoretical effort has emerged recently to calculate the third and higher moments of $P$ in various systems \cite{LevitovMath,LevRez,roche,Yuval,Nagaev,Kindermann,Kindermann2,Shelankov}. However, until now only the second moment has been measured in the many systems studied. 
Interest in the third and higher moments has occurred, first, because its character is predicted to differ significantly from that of the second moment; in particular, the third moment is insensitive to the sample 's own Johnson noise for the voltage bias case, yet is more sensitive, in a very subtle manner, to noise and loading by the environment.  Second, measurements of the higher moments may provide a new tool for studying conduction physics, complementary to the second moment.  It is the first set of issues that we address.

In this Letter we report the first measurements of the third moment of the voltage fluctuations across a conductor,  $\left<\delta V^3\right>$, where $\delta V(t)=V(t)-V$ represents the voltage fluctuations around the dc voltage $V$. Below we relate this to $\left<\delta I^3\right>$. Our experimental setup is such that the sample is current biased at dc and low frequency but the electromagnetic environment has an impedance $\sim50\;\Omega$ within the detection bandwidth, 10 MHz to 1.2 GHz. We have investigated tunnel junctions because they are predicted to be the simplest system having asymmetric current fluctuations. However, any kind of good conductor can be studied with the techniques we have developed. We studied two different samples, at liquid helium, liquid nitrogen and room temperatures. Our results are in agreement with a recent theory that considers the strong effect of the electromagnetic environment of the sample \cite{Kindermann2}. Moreover, we show that certain of these environmental effects can be dramatically reduced by signal propagation delays from the sample to the amplifier. This can guide future experiments on more exotic samples.

We present the theoretical overview first for the case of voltage bias. In a junction with a low transparency barrier biased by a dc voltage $V$, the current noise spectral density (related below to the second moment) is given by : $S_{I^2}=eGV \rm{coth}(eV/2k_BT)$ \cite{factor2} (in A$^2/$Hz), where $e$ is the electron charge and $G$ is the conductance. Only at high voltage $eV\gg k_BT$ does this reduce to the Poisson result $S_{I^2}=eI$ \cite{factor2}. If the barrier transparency is not small, the shot noise is $S_{I^2}=\eta eI$ where the Fano factor $\eta$ is smaller than 1 \cite{BuBlan}. For our samples $\eta\simeq1$. The total current fluctuations within a frequency bandwidth from $f_1$ to $f_2$ is given by $\left< \delta I^2\right>=2S_{I^2}(f_2-f_1)$. The spectral density of the third moment of the current fluctuations in a voltage biased tunnel junction of low transparency is calculated to be: $S_{I^3}=e^2GV$, independent of temperature \cite{LevitovMath,LevRez}. By considering how the Fourier components can combine to give a dc signal, we find that $\left<\delta I^3\right>=3S_{I^3}(f_2-2f_1)^2$. We have experimentally confirmed this unusual dependence on $f_1$ and $f_2$. 

We next consider the effects of the sample's electromagnetic environment (contacts, leads, amplifier, etc.); the sample is no longer voltage biased. The environment emits noise, inducing fluctuations of the voltage across the sample, which in turn modify the probability distribution $P$. Moreover, due to the finite impedance of the environment, the noise emitted by the sample itself induces also voltage fluctuations. This self-mixing of the noise by the sample is responsible for the Coulomb blockade of high resistance samples, where it induces a modification of the $I(V)$ characteristics \cite{Ingold}. We consider the circuit depicted in the inset of Fig. 2, at first neglecting time delay along the coaxial cable. The noise of the sample of resistance $R$ is modeled by a current generator $i$. The voltage $\delta V$ is measured across a resistor $R_0$, which has a current generator $i_0$ of noise spectral density $S_{i_0^2}$. One has $\delta V=-R_D(i+i_0)$ with $R_D=RR_0/(R+R_0)$ ($R$ in parallel with $R_0$). The spectral density of the second moment of the voltage fluctuations is $S_{V^2}=+R_D^2(S_{I^2}+S_{i_0^2})$. Thus, the only effect of the environment on the second moment of the noise is to rescale the fluctuations (by $R_D^2$) and add a bias-independent contribution \cite{BuBlan}.
In contrast, it has been recently predicted that the third moment of $P$ is significantly modified by the environment  \cite{Kindermann2}:
\begin{equation}
S_{V^3}=-R_D^3S_{I^3}+3R_D^4S_{i_0^2}\frac{dS_{I^2}}{dV}+3R_D^4S_{I^2}\frac{dS_{I^2}}{dV}
\end{equation}
The first term on the right is like that of the second moment. The negative sign results from an increasing sample current giving a reduced voltage. Our detection method is insensitive to $S_{i_0^3}$. The environment noise $i_0$  induces voltage fluctuations $\delta V=-R_Di_0$ across the sample. These modify the sample's noise $S_{I^2}$ (which depends on $V(t)$) as $-R_Di_0dS_{I^2}/dV$, to first order in $\delta V$ \cite{envS2}. This is the origin of the second term. The sample's own current fluctuations also modify the sample voltage to contribute similarly to the last term of Eq. (1).
We give below a simple derivation of how to include the effect of progation time in the coaxial cable, which dramatically affects $S_{V^3}$.

Two samples have been studied. Both are tunnel junctions made of Al/Al oxide/Al, using the double angle evaporation technique \cite{FultonDolan}. In sample A the bottom and top Al films are 50 nm thick. The bottom electrode was oxidized for 2 hours in pure $O_2$ at a pressure of 500 mTorr. The junction area is $15\;\mu$m$^2$. In sample B, the films are 120 nm and 300 nm thick, oxidation was for 10 min, and the junction area is $5.6\mu m^2$.
 
\begin{figure}
\includegraphics[width= 0.9\columnwidth]{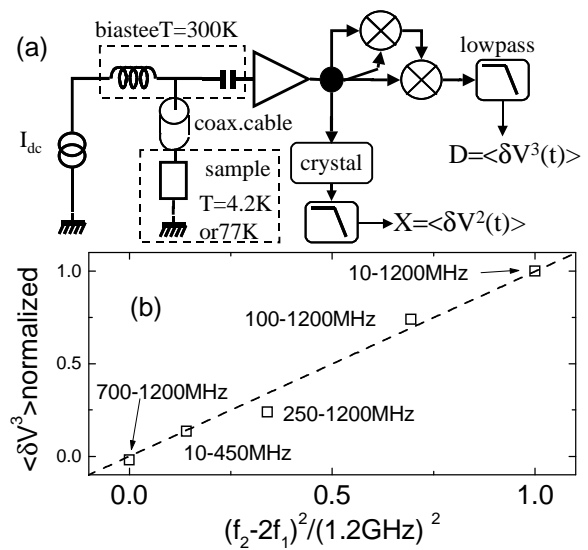}
\caption{(a) Schematic of the experimental setup. (b) Effect of finite bandwidth on the measurement of $\left<\delta V^3\right>$. Each point corresponds to a different value of the frequencies $f_1$ and $f_2$, as indicated in the plot. The data shown here correspond to sample B at $T=77$ K.}
\vspace{-3mm}
\end{figure}

We have measured $\delta V(t)^3$ in real time (see Fig. 1a).
The third moment of the voltage fluctuations $S_{V^3}$ is a very small quantity, and its measurement requires much more care and signal averaging than $S_{V^2}$ \cite{spurious}.   The sample is dc current biased through a bias tee. The noise emitted by the sample is coupled out to an rf amplifier through a capacitor so only the ac part of the current is amplified. The resistance of the sample is close to $50\;\Omega$, and thus is well matched to the coaxial cable and amplifier. After amplification at room temperature the signal is separated into four equal branches, each of which carries a signal proportionnal to $\delta V(t)$. A mixer multiplies two of the branches, giving $\delta V^2(t)$; a second mixer multiplies this result with another branch. The output of this second mixer, $\delta V^3(t)$, is then low pass filtered, to give a signal which we refer to as D. Ideally D is simply proportional to $S_{V^3}$, where the constant of proportionality depends on mixer gains and frequency bandwidth. The last branch is connected to a square-law crystal detector, which produces a voltage $X$ proportional to the the rf power it receives: the noise of the sample $\left<\delta V^2\right>$ plus the noise of the amplifiers. The dc current $I$ through the sample is swept slowly. We record $D(I)$ and $X(I)$; these are averaged numerically. This detection scheme has the advantage of the large bandwidth it provides ($\sim1$ GHz), which is crucial for the measurement. Due to the imperfections of the mixers, $D$ contains some contribution of $S_{V^2}$: $D=\alpha_3S_{V^3}+\alpha_2S_{V^2}$. We deduce $S_{V^3}=(D(I)-D(-I))/(2\alpha_3)$, since $S_{V^2}$ does not depend on the sign of $I$.

In order to determine the magnitude and sign of $\left<\delta V^3\right>$ we measured the signal D when the sample is replaced  by a programmable function generator. The output of the generator consist of a pseudo-random sequence of voltage steps (at a rate of $10^9$ samples per second, 1GS/s), the statistics of which can be specified. We measured the statistics with a 20GS/s oscilloscope.

\begin{figure}
\includegraphics[width= 0.9\columnwidth]{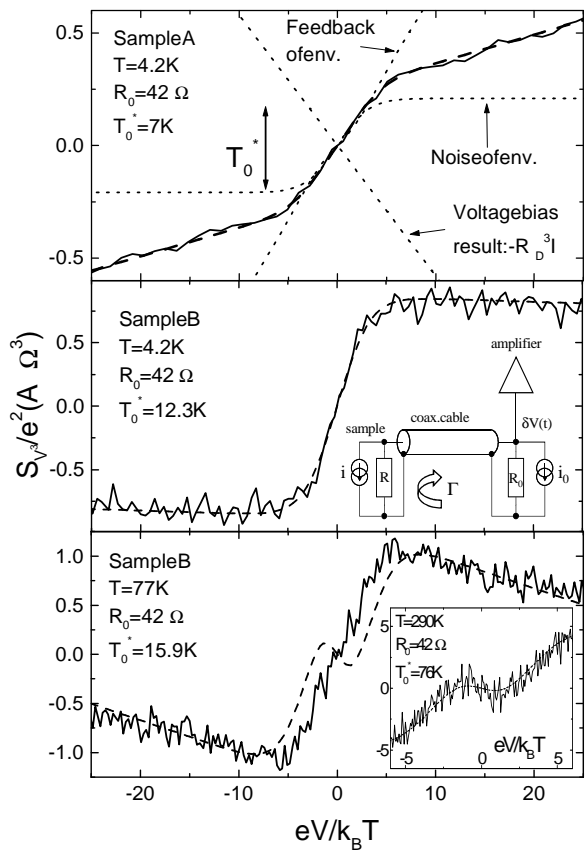}
\caption{Measurement of $S_{V^3}(eV/k_BT)$ (solid lines). The dashed lines corresponds to the best fit with Eq. (1). The dotted lines in the top plot correspond to the different contributions to $S_{V^3}$ (see text). Inset of the middle plot: schematics of the equivalent circuit used for the theoretical model.}
\vspace{-3mm}
\end{figure}

Sample A was measured at $T=4.2$K. Its total resistance (tunnel junction + contacts) is $62.6\;\Omega$. The resistance $R_A$ of the junction is extracted from the fit of $S_{V^2}$ as a function of $eV/k_BT$, with $V$ the voltage drop across the junction. We find  $R_A=49.6\;\Omega$. $R_{diff}$ is voltage independent to within  $1\%$. The gain of the amplification chain has been calibrated by replacing the sample by a macroscopic $50\;\Omega$ resistor whose temperature was varied. We find $\eta=1$ with a precision of a few percent for both samples. $S_{V^3}(eV/k_BT)$ for $|V|\leq10$ mV is shown in Fig. 2 (top); these data were averaged for 12 days.

Sample B was measured at $T=4.2$ K, 77 K and 290 K. The resistance of the junction $R_B=86\Omega$ is almost temperature independent. The contribution of the contacts is $\sim1\;\Omega$. In Fig. 2 (middle and bottom panels) the averaging time for each trace was 16 hours.

A powerful check that $D$ really measures $S_{V^3}$ is given by the effect of finite bandwidth. Each Fourier component $v(f)$ of $\delta V(t)$ is amplified by a frequency dependent gain $g(f)$ of the setup, such that $g(f)=1$ if $f_1<|f|<f_2$ and $g(f)=0$ otherwise. Here $f_1$ and $f_2$ are positive frequencies whereas $f$ has arbitrary sign. The measurement gives: $\left<\delta V^3\right>=\int \left<v(f)v(f')v(-f-f')\right>g(f)g(f')g(-f-f')=3S_{V^3}(f_2-2f_1)^2$ if $f_2>2f_1$, and 0 otherwise. A similar argument gives the well known result $\left<\delta V^2\right>=2S_{V^2}(f_2-f_1)$. The scaling of $S_{V^3}$ with $f_1$ and $f_2$ is characteristic of the measurement of a third order moment. We do not know any experimental artifact that has such behavior \cite{bandwidth}. $f_1$ and $f_2$ are varied by inserting filters before the splitter. As can be seen in Fig. 1b, our measurement follows the dependence on $(f_2-2f_1)^2$, which cannot be cast into a function of $(f_2-f_1)$ \cite{SV3bandwidth}.

To analyze our results, consider again the circuit in the inset of Fig. 2, a simplified equivalent of our setup. $R_0\sim50\;\Omega$ is the input impedance of the amplifier, which is connected to the sample through a coaxial cable of impedance $R_0$ (i.e., matched to the amplifier) . The sample's voltage reflection coefficient is $\Gamma=(R-R_0)/(R+R_0)$. In the analysis we present next we neglect the influence of the contact resistance and impedance mismatch of the amplifier, but we have included it when computing the theory to compare to the data. The voltage $\delta V(t)$ measured by the amplifier at time $t$  arises from three contributions: i) the noise emitted by the amplifier at time $t$: $R_0i_0(t)/2$ ; half of $i_0$ enters the cable. ii) the noise emitted by the sample (at time $t-\Delta t$, where $\Delta t$ is the propagation delay along the cable) that couples into the cable: $(1-\Gamma)Ri(t-\Delta t)/2$ ; iii) the noise emitted by the amplifier at time $t-2\Delta t$ that is reflected by the sample: $\Gamma R_0i_0(t-2\Delta t)/2$ ; thus,
\begin{equation}
\delta V(t)=-\frac{R_0}2\left[i_0(t)+\Gamma i_0(t-2\Delta t)\right]-\frac R2(1-\Gamma)i(t-\Delta t)
\end{equation}
For $\Delta t=0$, Eq. (2) reduces to $\delta V=-R_D(i+i_0)$ with $R_D=RR_0/(R+R_0)$. Thus, $\left<\delta V^3\right>=-R_D^3(\left<i^3\right>+3\left<i^2i_0\right>+3\left<ii_0^2\right>+\left<i_0^3\right>)$ for $\Delta t=0$. In this equation the term $\left<i^2i_0\right>$ leads to the second term on the right of Eq. (1). 
The term $\left<i^3\right>$ yields the first term of Eq. (1), and, due to the sample noise modulating its own voltage, the third term of Eq. (1) as well.
The terms $\left<i_0^3\right>$ and $\left<ii_0^2\right>$ are zero. The result for $\Delta t=0$ corresponds to Eq. (1), which is a particular case of Eq. (12b) of Ref. \onlinecite{Kindermann2}.

The finite propagation time does affect the correlator $\left<i^2i_0\right>$. The term $S_{i_0^2}$ in Eq. (1) has to be replaced by $(\Gamma S_{i_0^2}+S_{i_0(t)i_0(t-2\Delta t)})/(1+\Gamma)$, where $S_{i_0(t)i_0(t-2\Delta t)}$ is the spectral density corresponding to the correlator $\left<i_0(t)i_0(t-2\Delta t)\right>$. For long enough $\Delta t$ this term vanishes, since $i_0$ emitted at times $t$ and $t-2\Delta t$ are uncorrelated. Thus, the effect of the propagation time is to renormalize the noise temperature of the environment $T_0=R_0S_{i_0^2}/(2k_B)$ into $T_0^*=T_0\Gamma/(1+\Gamma)$.

\begin{figure}
\includegraphics[width= 0.9\columnwidth]{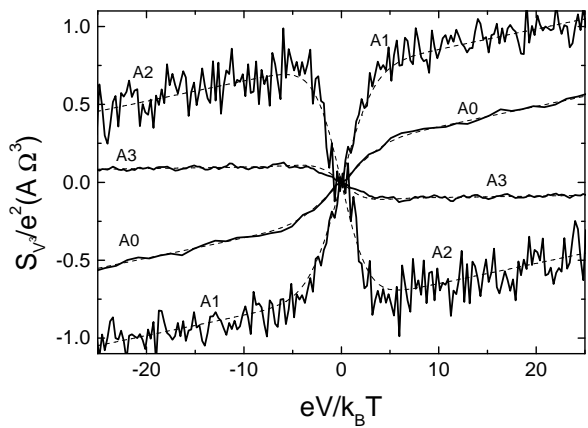}
\caption{Measurement of $S_{V^3}(eV/k_BT)$ for sample A at T=4.2 K (solid lines). A0: no ac excitation (same as Fig. 2, top panel). A1: with an ac excitation at frequency $\Omega/2\pi$ such that $\cos2\Omega\Delta t=+1$; A2: $\cos2\Omega\Delta t=-1$; A3: no ac excitation but a $63\;\Omega$ resistor in parallel with the sample. The dashed lines corresponds to fits with Eq. (1). }
\vspace{-3mm}
\end{figure}

We now check whether Eq. (1), with $S_{I^3}=e^2I$ and modified as above to account for finite propagation time, can explain our data. The unknown parameters are the resistance $R_0$ and the effective environment noise temperature $T_0^*$. We checked that the impedance of the samples was frequency independent up to 1.2 GHz within  5\%.  Fig. 2 shows the best fits to the theory, Eq. (1), for all our data. The four curves lead to $R_0=42\;\Omega$, in agreement with the fact that 
the electromagnetic environment (amplifier, bias tee, coaxial cable, sample holder) was identical for the two samples.
We have measured the impedance $Z_{env}$ seen by the sample. Due to impedance mismatch between the amplifier and the cable, there are standing waves along the cable. This causes $Z_{env}$ to be complex with a phase that varies with frequency. We measured that the modulus $|Z_{env}|$ varies between $30\;\Omega$ and $70\;\Omega$ within the detection bandwidth, in reasonable agreement with $R_0=42\;\Omega$ extracted from the fits.

We have measured directly the noise emitted by the room temperature amplifier; we find $T_0\sim100$ K. The cable of length $\sim2$m corresponds to $\Delta t$ long for the bandwidth we used. As a consequence, the relevant noise temperature to be used to explain the data is $T_0^*$. For sample A, $\Gamma=0.11$; including the contact resistance and cable attenuation one expects  $T_0^*=5$ K ; for sample B, $\Gamma=0.26$ and one expects $T_0^*=21$ K. A much shorter cable was used for $T=290$ K, and the reduction of $T_0$ is not complete. These numbers are in reasonably good agreement with the values of $T_0^*$ deduced from the fits (see fig. 2), and certainly agree with the trend seen for the two samples. Clearly $T_0^*\ll T_0$ for the long cable.

In order to demonstrate more explicitly the influence of the environment on $S_{V^3}$ we have conducted the following additional measurements. First, by adding a signal $A\sin \Omega t$ to $i_0$ (with $\Omega$ within the detection bandwidth) we have been able to modify $T_0^*$ without changing $R_0$, as shown on Fig. 3. The term $S_{i_0(t)i_0(t-2\Delta t)}$ oscillates like $\cos2\Omega\Delta t$, and thus one can enhance (curve A1 as compared to A0 in Fig. 3) or decrease $T_0^*$, and even make it negative (Fig. 3, A2). The curves A0--A2 are all parallel at high voltage, as expected, since the impedance of the environment remains unchanged; $R_0=42\;\Omega$ is the same for the fit of the three curves. Second, by adding a $63\;\Omega$ resistor in parallel with the sample (without the ac excitation) we have been able to change the resistance of the environment $R_0$, and thus the high voltage slope of $S_{V^3}$. The fit of curve A3 gives $R_0=24.8\;\Omega$, in excellent agreement with the expected value of $24.7\;\Omega$ ($63\;\Omega$ in parallel with $42\;\Omega$). Since this makes the reflection coefficient negative ($\Gamma=-0.22$), the presence of the extra resistor also reverses the sign of $T_0^*$, as expected.  

Our data are consistent with a third moment of current fluctuations $S_{I^3}$ being independent of $T$ between 4K and 300K when the sample is voltage biased, as predicted for a tunnel junction. The effect of the environment, through its noise and impedance, is clearly demonstrated. This is of prime importance for designing future measurements on samples with unknown third moment.

We thank C. Wilson and L. Spietz for providing samples A and B, respectively. We thank C. Beenakker, W. Belzig, A. Clerck, M.~Devoret, Y. Gefen, S. Girvin, N. Hengartner, M. Kindermann, L. Levitov, A. Mukherjee, Y. Nazarov and P.-E. Roche for useful discussions. This work was supported by NSF DMR 0072022.


\begin{thebibliography}{99}
\bibitem{envS2} There are terms of order $\delta V^2$ and higher, that are small and neglected here. Such terms affect slightly $S_{V^2}$ as well.
\bibitem{spurious} We have checked various potential problems: the non linearities of the amplifiers, the imperfections of the mixers, the effect of change of the bias point of the mixers due to the noise of the sample, the noise coming from the dc source, the time delay between the branches feeding the various mixers, and the asymmetry of $S_2$ vs. current.
\bibitem{bandwidth} Each point on the curve of Fig. 1b corresponds to a full $\left<\delta V^3(I)\right>$ measurement. An asymmetric external noise would just add just a constant to $D$, and thus would not affect our measurement, even though this constant might have the same dependence on bandwidth as $\left<\delta V^3\right>$.
\bibitem{SV3bandwidth} Both $S_{I^3}$ and $S_{V^3}$ have the same dependence on bandwidth, if the environment impedance is constant.
\end{thebibliography}
\end{document}